\documentstyle[preprint,aps,epsfig]{revtex}

\begin{document}

\draft

\title{Alpha-decay Rates of Yb and Gd in Solar Neutrino Detectors}

\author{M.~Fujiwara,$^{1,2}$ T.~Kawabata,$^3$ P.~Mohr$^4$}

\address{
$^1$ Research Center for Nuclear Physics, Osaka University,
       Mihogaoka 10-1, Ibaraki, Osaka 567-0047, Japan \\
$^{2}$Advanced Science Research Center, JAERI, Tokai, Ibaraki,
       319-1195 Japan \\
   $^3$ Department of Physics, Kyoto University,
       Kyoto 606-8502, Japan \\
  $^4$ Institut f\"ur Kernphysik, Technische Universit\"at Darmstadt,
       Schlossgartenstrasse 9, D--64289 Darmstadt, Germany
}
\date{\today}
\maketitle
\begin{abstract}
The $\alpha$-decay rates for the nuclides
$^{168,170,171,172,173,174,176}$Yb and $^{148,150,152,154}$Gd have
been estimated from transmission probabilities in a systematic
$\alpha$-nucleus potential and from an improved fit to $\alpha$-decay
rates in the rare-earth mass region. Whereas ${\alpha}$-decay of 
$^{152}$Gd in natural gadolinium is a severe obstacle for the use of
gadolinium as a low-energy solar-neutrino detector, we show that
 ${\alpha}$-decay does not contribute significantly to the background
in a ytterbium detector. 
An extremely long ${\alpha}$-decay lifetime of $^{168}$Yb is
obtained from calculation, which may be close to the sensitivity limit in a
low-background solar neutrino detector.
\end{abstract}


% 23.60.+e  alpha decay
% 26.65.+t  Solar neutrinos
% 27.60.+j  90 < A < 149
% 27.70.+q  150 < A <  189

% \begin{multicols}{2}
\narrowtext

  The solar neutrino problem is an important current subject to be
studied in relation to the fundamental physics of neutrino oscillations.
After the successful measurements of the solar neutrinos by
Super-Kamiokande\cite{fukuda}, SAGE\cite{Sage}, GALLEX\cite{Gallex},
and SNO\cite{Sno}, the problem appears to be solved best in terms of
oscillation of ${\nu}_e$ into other neutrino flavors. To arrive at the
final solution of the solar-${\nu}$ problem, a real-time measurement
of the ${\nu}_e$-energy spectrum including $pp$-, $^7$Be-neutrinos 
is now of central interest \cite{Bahcall00}. 

The solar neutrino detection requires an extremely low-background
measurement with new technologies. Raghavan \cite{raju} has suggested
a flavor-specific detection scheme with low thresholds for the
real-time detection of solar neutrinos. The neutrino captures 
$\nu_e + ^{176}{\rm{Yb}} \rightarrow e^- + ^{176}{\rm{Lu}}^\ast$ and
$\nu_e + ^{160}{\rm{Gd}} \rightarrow e^- + ^{160}{\rm{Tb}}^\ast$ are
based on charged current mediated Gamow-Teller transitions. For the
identification of a neutrino capture event a delayed coincidence
between a prompt $e^-$ and a $\gamma$-ray from the decay of a
nanosecond isomer in $^{176}$Lu or $^{160}$Tb is used. For both
suggested nuclei, $^{176}$Yb and $^{160}$Gd, one finds sufficient
Gamow-Teller strength at low energies and a nanosecond isomer which is
populated in the neutrino capture reaction. Typical detection rates of
solar neutrinos are of the order of ten counts per year and per ton of
detector material \cite{Fuji2000}.

A new solar neutrino detector using Yb materials has been proposed in
LGNS at Gran Sasso \cite{raju,Fuji2000,memo}, and sources of background for
such a detector are presently being studied \cite{Meyer01}.
The ${\alpha}$-decay from the material
in a solar-${\nu}$ detector is a possible serious problem to pollute 
the spectrum consisting of rare neutrino events. For example, it has
turned out that Gd material is not suitable for a solar-${\nu}$ detector;
the $^{152}$Gd isotope decays by ${\alpha}$ emission with a half-life 
of $T_{1/2}=1.08{\times}10^{14}$ years, and is followed by the 
subsequent $^{148}$Sm ${\alpha}$-decay with $T_{1/2}=7{\times}10^{15}$ years.
The number of $^{152}$Gd atoms included in the natural Gd metal with
20 tons amounts to $1.5{\times}10^{26}$ atoms. The decay rate 
calculated by the well-known radioactive decay formula
\begin{equation}
 \frac{dN}{dt} = -N_0\lambda \, e^{-{\lambda}t}
\end{equation}
amounts to about 30,000 counts/sec for the 20 ton Gd material. This is a huge
decay rate, which decreases the reliability of the coincidence
events due to a large number of background ${\alpha}$-signals and due
to accidental coincidences even in the case of a narrow time gate
less than $10^{-8}$ seconds for coincidence.

Recently, ytterbium has been suggested as material for a solar-$\nu$
detector. The Q-values of
${\alpha}$-particle emission for all the Yb isotopes are 
positive, but relatively small. Hence, each Yb isotope
must have an extremely  long half-life although 
there has been no report on the
${\alpha}$-decay lifetime in literature. Such
half-lifes may be measurable
with a large mass detector like a solar-${\nu}$ detector.
If the half-life of some Yb isotope is relatively short, the
${\alpha}$-decay
may cause a significant background in the analysis of the obtained
neutrino data.
Therefore, it is important to estimate the ${\alpha}$-decay rates of
the Yb isotopes and to compare them with the solar neutrino
rates. 
Recently, Gamow-Teller strengths
in the inverse $\beta$ transition $^{176}$Yb $\rightarrow$ $^{176}$Lu
were studied, and it has been concluded that Yb-based
detectors are well-suited for the real-time spectroscopy of the sub-MeV solar
neutrinos \cite{Fuji2000,Bha2000}.

There are several parameterizations for $\alpha$-decay rates available
mainly for nuclei in the region of $Z \geq 84$ \cite{ABrown,Taage,Keller}.  
The predictions from these
parameterizations usually show small deviations for
experimentally available decay rates. However,
there often happen large discrepancies, up to ten orders of
magnitude, between the
predictions for extremely small decay rates (corresponding to small
$Q_\alpha$ values). In this report, we wish to discuss the
${\alpha}$-decay rates of Yb and Gd isotopes in the rare earth nuclei.

The lifetime of ${\alpha}$-decay was first formulated by Geiger
and Nuttall \cite{Geiger}.
The probability of ${\alpha}$-particle emission, $W$, can be expressed as
\begin{equation}
 W = p_{\alpha}{\nu}T,
\end{equation}
where $p_{\alpha}$ is the probability of finding an
${\alpha}$-particle in a nucleus, ${\nu}$ the frequency of
an ${\alpha}$-particle appearing at the wall of a
nucleus, and $T$ is the transmission coefficient for the tunnelling effect.
The half-life  $T_{1/2}$ can be parameterized \cite{Wong} as
\begin{equation}
 log_{10}T_{1/2} = a_{1} + log_{10}\frac{\sqrt{Q_{\alpha}}}{A^{1/3}} +
 a_2 {\sqrt{Z_dA^{1/3}}} + a_3\frac{Z_d}{\sqrt{Q_{\alpha}}} \quad \quad,
\label{eq:wong}
\end{equation}
with the half-life $T_{1/2}$ in years, the $Q$-value for $\alpha$-decay
$Q_\alpha$ in MeV, and the mass (proton, neutron) numbers $A$ ($Z,N$)
of the parent nucleus and $A_d = A-4$ ($Z_d = Z-2$, $N_d = N-2$) of
the daughter nucleus.

\begin{figure}
\begin{center}
\epsfig{figure=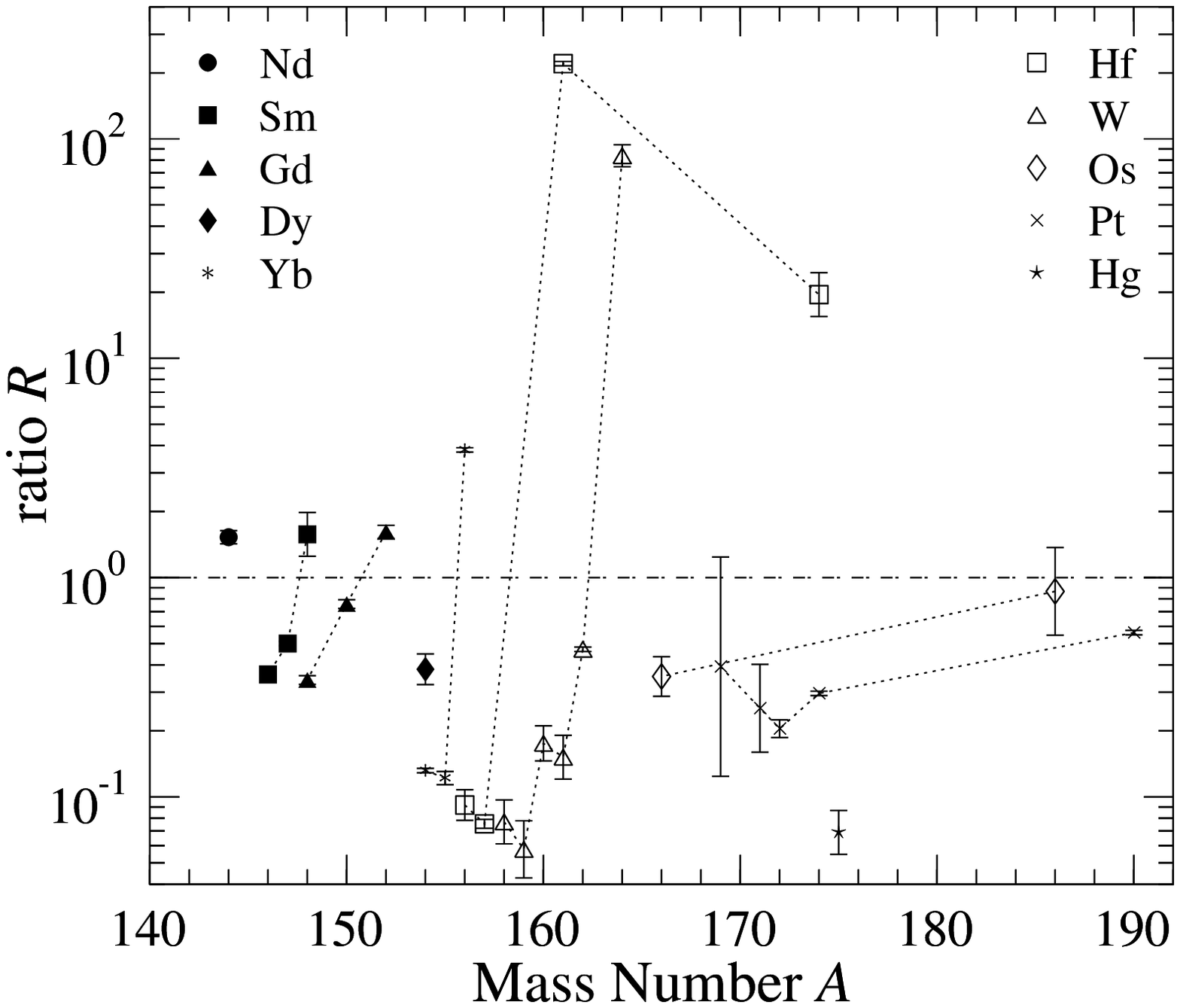,bbllx=69,bblly=66,bburx=527,bbury=457,width=12.0cm}
\end{center}
\caption{
  \label{fig:fit}
	Ratio $R = T_{1/2}^{\rm{calc}}/T_{1/2}^{\rm{exp}}$
	between the calculated half-lives $T_{1/2}^{\rm{calc}}$
	with Eq.~(\ref{eq:wong}) to the experimental half-lives 
	$T_{1/2}^{\rm{exp}}$ for nuclei in the rare
	earth region. The dotted lines connect isotopes from the same
	element. The error bars are only experimental. Additional
	errors from the uncertainty of $Q_\alpha$ are not included
	(see last lines of Tables \ref{Yb-a} and \ref{Gd-a}).
	The analyzed range of $Q$-values and half-lives is broad:
	$Q$-values cover the range from 1.9\,MeV ($^{144}$Nd) to
	7.0\,MeV ($^{175}$Hg), and half-lives $T_{1/2}$ cover 26
	orders of magnitude from milliseconds ($^{158}$W, $^{169}$Pt)
	to more than $10^{15}$ years ($^{144}$Nd, $^{148}$Sm,
	$^{174}$Hf, $^{186}$Os).
}
\end{figure}

Since the data fittings in most of Refs.~\cite{ABrown,Taage,Keller}
are focused on the ${\alpha}$-decay for
the nuclei with $Z>80$, we tried to fit the ${\alpha}$-decay
half-lives for nuclei in the rare earth region separately.
The data for the half-lives of the ${\alpha}$-decay
nuclei in the rare earth region were fitted using
Eq.~(\ref{eq:wong}) with $a_1$, $a_2$, and
$a_3$ as three parameters. 
All $Q_\alpha$ values have been taken from the
mass table of Audi and Wapstra \cite{Audi}.
 The best fitting parameters, $a_1$, $a_2$, and
$a_3$ are $a_1=-34.577$, $a_2=-1.2685$, and $a_3=1.7336$, respectively.
The $\alpha$-decay half-lives are found to be well fitted
as can be seen from
Fig.~\ref{fig:fit}.
The $\chi^2$ per degree of freedom has been significantly improved. 
In our 3-parameter fit $\chi^2$ is reduced by a factor of about 4
compared to the 2-parameter fit of Ref.~\cite{ABrown}, and by a factor
of about 2 compared to the 2-parameter fit of
Refs.~\cite{Taage,Keller} which is based
on the model of Bethe \cite{Bethe37}. These 2-parameter fits
are given by \cite{ABrown}
\begin{equation}
log_{10}T_{1/2} = (9.54Z_d^{0.6}/\sqrt{Q_{\alpha}}) - 58.87
\label{eq:ABrown}
\end{equation}
and \cite{Taage,Keller}
\begin{equation}
log_{10}T_{1/2} = 1.598(Z_d/\sqrt{Q_{\alpha}} - Z_d^{2/3})- 27.44
\label{eq:taagekeller}
\end{equation}

Tables \ref{Yb-a} and \ref{Gd-a} show the expected half-lives for the
Yb and Gd isotopes.
The experimental half-life of $1.1{\times}10^{14}$ years
for $^{152}$Gd agrees with the result of the present fitting
within a factor of 1.4. However, the predictions from the two
parameter fittings by Brown \cite{ABrown} 
and Refs.~\cite{Taage,Keller}
differ by three orders of magnitude. A similar agreement is also found
for $^{148}$Gd and $^{150}$Gd.  Again there are strong deviations
between the predictions mainly given from the fits of ${\alpha}$-decay
rates for nuclei with $Z \geq 84$ \cite{ABrown,Taage,Keller}. Here, one can
address the question that the descrepancy may come from the
deformation effect on ${\alpha}$-decay for the rare earth nuclei. But,
it should be noted that the most nuclei with $Z > 84 $ are deformed as
well, and the influence of nuclear deformation deos not change the
half-life by an order of magnitude (see discussion below).

Our prediction for the different quasi-stable Yb isotopes varies from
$T_{1/2} = 4.5{\times}10^{24}$ years for $^{168}$Yb to
$T_{1/2} = 6.2{\times}10^{95}$ years for $^{176}$Yb.
The parameterizations from literatures are between one order of
magnitude ($^{168}$Yb) smaller \cite{Taage,Keller} (larger
\cite{ABrown}) to five orders of magnitude ($^{176}$Yb) smaller
\cite{Taage,Keller} (larger \cite{ABrown}). Note that no useful
prediction for the relevant $\alpha$-decay half-life of $^{168}$Yb can
be derived from Refs.~\cite{ABrown,Taage,Keller} because their
predictions differ by more than three orders of magnitude.
Of course, the calculated half-lives of $^{172,173,174,176}$Yb exceed
any measurable range. But, the excellent agreement between the new fit
and the half-lives from the folding potential calculations (see below)
gives us
some confidence that our predictions are reliable in a very broad range of
half-lives (where other extrapolations show huge deviations).

Because of these large discrepancies between the extrapolations of the
various parameterizations, we  applied the semi-classical
model of Gurvitz and K\"albermann \cite{Gur87} together with the
systematic folding potentials of Atzrott {\it et al.}
\cite{Atz96}. The required nuclear densities were derived from
measured charge density distributions \cite{deVries87}. When no measured
charge distribution was available, the measured distribution of a
neighboring isotope was used with an adjusted radius parameter $R \sim
A^{1/3}$. This model was already successfully used to calculate the
$\alpha$-decay properties of neutron-deficient so-called $p$-nuclei in
\cite{Mohr2000}.

In the semi-classical model, the $\alpha$-decay width
$\Gamma_\alpha$ is given as \cite{Gur87}:
\begin{equation}
\Gamma_\alpha = PF\frac{\hbar^2}{4\mu}
\exp{\left[ -2 \int_{r_2}^{r_3} k(r) dr \right]}
\label{eq:gamma}
\end{equation}
with the preformation factor $P$, the normalization factor $F$
\begin{equation}
F \int_{r_1}^{r_2} \frac{dr}{k(r)} = 1
\label{eq:f}
\end{equation}
and the wave number $k(r)$
\begin{equation}
k(r) = \sqrt{ \frac{2\mu}{\hbar^2}\left|Q_\alpha - V(r)\right|} 
\quad \quad ,
\label{eq:k}
\end{equation}
where $\mu$, $Q_\alpha$, and  $r_i$  are the reduced mass,
the $\alpha$-decay energy, and the classical turning points,
respectively. 
For $0^+ \rightarrow 0^+$
$s$-wave decay, the inner turning point is at $r_1 = 0$. $r_2$ varies
from about 7 to 9 fm, and $r_3$ varies up to about 350 fm for the
lowest $Q_\alpha$ value.
The decay width $\Gamma_\alpha$ is related
to the half-life by the well-known relation
$\Gamma_\alpha = \hbar \ln{2} / T_{1/2}$.
For the odd Yb isotopes, an additional centrifugal potential was used.
The preformation factors $P$ were estimated from the systematic trend
of $P$ shown in Fig.~2 of Ref.~\cite{Mohr2000} leading to $P =
(4.0 \pm 0.8)$\,\% for the Yb isotopes and $P = (10.4 \pm 2.5)$\,\% for the Gd
isotopes. Larger $P$ values can be expected around the shell closures
$N=82$ and $N=126$, whereas smaller values for $P$ are found in between.
The results of this model are listed in  Tables
\ref{Yb-a} and \ref{Gd-a}. They are in  good agreement with our new fits
in the rare earth region. Therefore, the newly
determined $\alpha$-decay formula can be used for predictions in the
rare earth region with very limited uncertainties.

A systematic study of $\alpha$-decay half-lives \cite{Buck93} using
the same model \cite{Gur87} but specially shaped potentials is in rough
agreement with our results for the Gd isotopes (8$^{\rm{th}}$ line in Table
\ref{Gd-a}). Unfortunately, the
quasi-stable Yb isotopes are not considered in Ref.~\cite{Buck93}.

There are several sources of uncertainties for the presented
predictions for the half-lives of the Yb and Gd isotopes.
Because of the roughly exponential dependence of the tunneling
probability on the $\alpha$ decay energy $Q_\alpha$, a factor
$f(\Delta Q_\alpha)$ has been calculated by which the half-lives are
shorted (enlarged) if $Q_\alpha$ is increased (decreased) by its
uncertainty $\Delta Q_\alpha$. This factor $f(\Delta Q_\alpha)$ is
given in Tables \ref{Yb-a} and \ref{Gd-a}. Typical uncertainties
remain within 40\,\% with the exception of $^{176}$Yb where the very
low $Q_\alpha$ and its 4\,keV uncertainty lead to an uncertainty of
$f(\Delta Q_\alpha) = 3.61$. This uncertainty is common to all
predictions in this work and also in Refs.~\cite{ABrown,Taage,Keller}.

The uncertainties of the predictions from the 3-parameter fit in this
work are smaller than in the previous 2-parameter fits because of the
improved $\chi^2$ value. The average deviation for $log_{10}T_{1/2}$
in our fit is about 0.5 corresponding to an average deviation of a
factor of 3. Therefore, a typical uncertainty of a factor of 3 can be
expected for our predicted half-lives. The uncertainties in the
coefficients $a_1$, $a_2$, and $a_3$ lead to minor unertainties
compared to the average deviation of a factor of 3.

In the case of the folding potential calculations,  the predicted values
depend sensitively on the average preformation factors derived from
the systematics in Ref.~\cite{Mohr2000}. Typical uncertainties of $P$ are
of the order of $20-30$\,\% which lead to identical uncertainties for
the predicted half-lives. Together with the uncertainties of
$Q_\alpha$, this leads to an overall uncertainty of better than a
factor of two for the folding potential predictions.

It is well-known that nuclei in the rare earth region show strong
deformations with typical deformation parameters in the order of
$\beta_2 \approx 0.3$ \cite{Raman87} which may lead to additional
uncertainties. The influence of deformation on
the decay probabilities has been analyzed recently in
Refs.~\cite{Ste96,Del97,Gar00,Dim00}. In Ref.~\cite{Ste96}, it is
stated that various calculations of the half-lives agree within a
factor of 4 with the 
spherical result. In Ref.~\cite{Del97}, it is suggested that 
the deformation effect 
enhances the decay probability by roughly a factor of 2, whereas
in Ref.~\cite{Gar00} the decay probability for deformed nuclei is reduced
by a factor less than 2. In Ref.~\cite{Dim00}, the deformation
effect leads only to
small variations of the half-lives.
The good quality of our fit to the decay
data in the rare earth region confirms that the influence of
deformation on the $\alpha$-decay half-lives remains relatively small
compared to the uncertainties that arise from the predictions of
our and of previous fits \cite{ABrown,Taage,Keller}.

In summary, we estimated the half-lives of ${\alpha}$-decay for the Gd
and Yb isotopes. Our predictions from the new fit in the rare-earth
region and from the folding potential calculations agree within their
uncertainties discussed above, whereas previous fits show large
deviations especially for low $Q_\alpha$ values and long half-lives.
The ``short'' half-life of $^{152}$Gd of 
$T_{1/2} = 1.1{\times}10^{14}$ years  leads to serious 
background problems in a
neutrino detector. The estimated half-life for $^{168}$Yb is between
$4.5{\times}10^{24}$ years and $1.1{\times}10^{25}$ years. If this
predicted half-life is correct, the LENS detector with a mass of 20
tons of natural Yb will detect about
6${\sim}$14 counts/year for ${\alpha}$-particle emission with
$Q_{\alpha}$=1.951 MeV from $^{168}$Yb decay, being negligble compared
with the solar-${\nu}$ events. This low rate causes
definitely no serious background problem.  However,
the longest ${\alpha}$-decay
radioactivity so far measured in the past may be found in a
low-background detector designed for a real-time solar-${\nu}$  spectroscopy.

%%%%%%%%%%% Figures %%%%%%%%%%%

%%%%%%%%%%% Tables %%%%%%%%%%%

\begin{table}
\caption{Isotopic abundance (\%) of Yb,
${\alpha}$-decay energy ($Q_{\alpha}$),
the expected half-lives ($T_{1/2}$) in unit of year,
and the uncertainty factor $f(\Delta Q_\alpha)$ from the $Q$ value error. 
The parameters of Eq.~(\ref{eq:wong}) were adjusted to the half-lives
of the nuclides shown in Fig.~\ref{fig:fit}.
}
\begin{center}
\begin{tabular}{cccccccc}
\hline
& $^{168}$Yb & $^{170}$Yb  & $^{171}$Yb &  $^{172}$Yb
	&  $^{173}$Yb & $^{174}$Yb & $^{176}$Yb \\
\hline
   (\%) &  0.13 & 3.05 & 14.3 & 21.9 & 16.1 & 31.8 & 12.7 \\
$Q_{\alpha}$ (MeV)  
	& 1.95075 	& 1.73764 	& 1.55895
	& 1.31030 	& 0.94586	& 0.74009	& 0.57102 \\
$\Delta Q_{\alpha}$ (MeV)  
	& 0.00412 	& 0.00146	& 0.00139
	& 0.00139	& 0.00145	& 0.00160 	& 0.00417 \\
$T_{1/2}$ (exp.) & -     & -  & -    &  -   & -   & - & -   \\
$T_{1/2}$(est.) \tablenotemark[1]
	& $1.1{\cdot}10^{27}$  & $1.4{\cdot}10^{32}$
	& $1.6{\cdot}10^{37}$  & $8.5{\cdot}10^{45}$
	& $3.0{\cdot}10^{64}$  & $3.8{\cdot}10^{80}$
	& $7.7{\cdot}10^{99}$  \\
$T_{1/2}$ (est.) \tablenotemark[2]
	& $5.5{\cdot}10^{23}$  & $2.4{\cdot}10^{28}$
	& $9.3{\cdot}10^{32}$  & $7.4{\cdot}10^{40}$
	& $4.7{\cdot}10^{57}$  & $1.8{\cdot}10^{72}$
	& $5.5{\cdot}10^{89}$      \\
$T_{1/2}$ (est.) \tablenotemark[3]
	& $4.5{\cdot}10^{24}$  & $4.1{\cdot}10^{29}$
	& $3.5{\cdot}10^{34}$  & $1.1{\cdot}10^{43}$
	& $1.5{\cdot}10^{61}$  & $8.4{\cdot}10^{76}$
	& $6.2{\cdot}10^{95}$     \\
$T_{1/2}$ (est.) \tablenotemark[4]
	& $1.1{\cdot}10^{25}$  & $6.9{\cdot}10^{29}$
	& $4.7{\cdot}10^{35}$  & $1.0{\cdot}10^{43}$
	& $2.1{\cdot}10^{61}$  & $1.9{\cdot}10^{76}$
	& $6.9{\cdot}10^{94}$     \\
$f(\Delta Q_\alpha)$
	& 1.24	& 1.09	& 1.10	& 1.13	& 1.23	& 1.40	& 3.61 \\
\hline
\end{tabular}
\end{center}
\tablenotetext[1]{Parametrization of B.~A.~Brown \cite{ABrown}.}
\tablenotetext[2]{Parametrization of R.~Taagepera and 
M.~Nurmia \cite{Taage},
	K.~K.~Keller and H.~Z.~Munzel \cite{Keller}.}
\tablenotetext[3]{this work, from fit to rare earth region.}
\tablenotetext[4]{this work, from folding potential,
	using $P = (4.0 \pm 0.8)\,\%$.}
\label{Yb-a}
\end{table}

\begin{table}
\caption{Isotopic abundance (\%) of Gd,
${\alpha}$-decay energy ($Q_{\alpha}$), the
expected half-lives ($T_{1/2}$) in unit of year,
and the uncertainty factor $f(\Delta Q_\alpha)$ from the $Q$ value error. 
The parameters of Eq.~(\ref{eq:wong}) were adjusted to the half-lives 
of the nuclides shown in Fig.~\ref{fig:fit} including $^{148,150,152}$Gd.
}
\begin{center}
\begin{tabular}{ccccc}
\hline
  & $^{148}$Gd & $^{150}$Gd & $^{152}$Gd & $^{154}$Gd  \\
\hline
  (\%)  & - & - &  0.20 & 2.18 \\
$Q_{\alpha}$ (MeV)
	& 3.27121 & 2.80887  & 2.20458 & 0.91991 \\
$\Delta Q_{\alpha}$ (MeV)
	& 0.00003 & 0.00630  & 0.00142 & 0.00115 \\
$T_{1/2}$ (exp.)
	& $7.5{\cdot}10^{1}$	& $1.8{\cdot}10^{6}$
	& $1.1{\cdot}10^{14}$ 	& -  \\
$T_{1/2}$ (est.) \tablenotemark[1]
	& $7.6{\cdot}10^{3}$	& $7.1{\cdot}10^{8}$
	& $3.7{\cdot}10^{17}$	& $2.9{\cdot}10^{59}$ \\
$T_{1/2}$ (est.) \tablenotemark[2]
	& $2.0{\cdot}10^{2}$	& $4.4{\cdot}10^{6}$
	& $1.8{\cdot}10^{14}$	& $6.7{\cdot}10^{50}$ \\
$T_{1/2}$ (est.) \tablenotemark[3]
	& $2.6{\cdot}10^{1}$	& $1.1{\cdot}10^{6}$
	& $1.5{\cdot}10^{14}$	& $4.1{\cdot}10^{53}$ \\
$T_{1/2}$ (est.) \tablenotemark[4]
	& $6.0{\cdot}10^{1}$	& $1.7{\cdot}10^{6}$
	& $1.4{\cdot}10^{14}$	& $4.1{\cdot}10^{52}$ \\
$T_{1/2}$ (est.) \tablenotemark[5]
	& $2.6{\cdot}10^{1}$	& $9.0{\cdot}10^{5}$
	& $6.7{\cdot}10^{13}$	& - \\
$f(\Delta Q_\alpha)$
	& 1.00	& 1.18	& 1.06	& 1.17 \\
\hline
\end{tabular}
\end{center}
\tablenotetext[1]{Parameterization of B.~A.~Brown \cite{ABrown}.}
\tablenotetext[2]{Parameterization of R.~Taagepera and 
M.~Nurmia \cite{Taage},
	K.~K.~Keller and H.~Z.~Munzel \cite{Keller}.}
\tablenotetext[3]{this work, from fit to rare earth region.}
\tablenotetext[4]{this work, from folding potential,
	using $P = (10.4 \pm 2.5)\,\%$.}
\tablenotetext[5]{from Ref.~\cite{Buck93}.}
\label{Gd-a}
\end{table}

% \end{multicols}

\end{document}